\documentclass[twoside,slac_one]{revtex4}
\usepackage{graphicx}
\usepackage{fancyhdr}
\usepackage{amsmath} % American Mathematics Society standards
\usepackage{bm}% bold math
\usepackage{amsxtra}
\usepackage{amssymb}
\usepackage{amsthm}
\usepackage{latexsym}
\usepackage{lscape}

\begin{document}

\title{Measurement of the Relative Branching Fraction
of \boldmath $B_{s}^{0} \rightarrow J/\psi f_{0}(980), f_{0}(980) \rightarrow \pi^{+}\pi^{-}$ to $B_{s}^{0} \rightarrow J/\psi \phi, \phi \rightarrow K^{+}K^{-}$}


\author{B. Abbott}
\affiliation{Homer L. Dodge Department of Physics and Astronomy, University of Oklahoma, Norman, OK, USA}

\begin{abstract}
A measurement of the relative branching fraction of $B_{s}^{0} \rightarrow J/\psi f_{0}(980), 
f_{0}(980) \rightarrow \pi^{+}\pi^{-}$ to $B_{s}^{0} \rightarrow J/\psi \phi, \phi \rightarrow K^{+}K^{-}$ is presented.
The decay mode $B_{s}^{0} \rightarrow J/\psi f_{0}(980)$ is an interesting mode since it is a  
CP-odd eigenstate which could be used in CP-violating studies.  Using approximately
8~$\rm{fb}^{-1}$ of data recorded with the D0 detector at the Fermilab Tevatron Collider, a
relative branching fraction of 0.210 $\pm$ 0.032\thinspace(stat) $\pm$ 0.036\thinspace(syst) is found.

\end{abstract}

\maketitle

\thispagestyle{fancy}


%%%%%%%%%%%%%%%%%%%%%%%%%%%%%%%%%%
\section{Introduction}

The CP-violating phase in $B_{s}^{0}$ mixing  mixing,
has been measured \cite{d0phis,cdfphis} using  $B_{s}^{0} \rightarrow J/\psi \phi$ decays. 
The measured absolute value is larger than predicted by the Standard Model (SM) \cite{SMphi}, but is 
statistically consistent with it.  The decay products in $B^0_s \rightarrow J/\psi f_{0}(980)$ are in a CP-odd eigenstate
and can provide a more direct measurement of this CP-violating phase.
Measuring this CP-violating phase using  $B^0_s \rightarrow J/\psi f_{0}(980)$ decays mode can aid in reducing its uncertainty.

Based on estimates the relative branching fraction should be large.  Using hadronic $D^{+}_{s}$ decays, Stone and 
Zhang \cite{Stone,StoneZhang} estimated the relative width to be:

\begin{equation}
R \equiv \frac { \Gamma (B_{s}^{0}\rightarrow J/\psi f_{0}(980);f_{0}(980)\rightarrow \pi^{+} \pi^{-}) } 
                  { \Gamma (B_{s}^{0}\rightarrow J/\psi \phi;\phi \rightarrow K^{+} K^{-}    )} \approx 0.20. 
\end{equation}

The LHCb collaboration has reported \cite{LHCb} 
a first measurement of $R=0.252^{+0.046 +0.027}_{-0.032 -0.033}$. 
The Belle collaboration has made a measurement of the branching
fraction $\mathcal{B} (B_{s}^{0}\rightarrow J/\psi  f_{0}(980); 
 f_{0}(980) \rightarrow \pi^{+} \pi^{-})$ = 
$(1.16^{+0.31}_{-0.19}\thinspace(\rm{stat.})^{+0.15}_{-0.17}\thinspace(\rm{syst.})^{+0.26}_{-0.18}(N_{B_{s}^{(*)}\bar{B}_{s}^{(*)}})) 
\times 10^{-4}$ \cite{Bellemeasure}.
The CDF collaboration has also measured the relative branching fraction and finds $R$=0.257 $\pm$ 0.020\thinspace(stat) $\pm$ 0.014\thinspace(syst) 
\cite{CDFf0}.
This article provides a new measurement of the relative branching fraction using the D0 detector collecting data at the Fermilab
Tevatron Collider.

This note provides a new measurement of the relative branching fraction from D0.

\section{Relative Branching Fraction}

To determine an absolute branching fraction, various efficiencies, branching fractions,
and cross sections need to be known, as well as the integrated luminosity. However, by measuring a relative 
branching fraction, several terms common to both the $B_{s}^{0}\rightarrow J/\psi f_{0}(980)$ branching
fraction and the $B_{s}^{0}\rightarrow J/\psi \phi$ branching fraction cancel giving:
\begin{equation}
R =  \frac{ \mathcal{B} (B_{s}^{0}\rightarrow J/\psi f_{0}(980)    ;f_{0}(980)    \rightarrow \pi^{+} \pi^{-})  }
        { \mathcal{B} (B_{s}^{0}\rightarrow J/\psi \phi;\phi \rightarrow K^{+} K^{-}    )} = 
\frac{N_{ B_{s}^{0} \rightarrow J/\psi f_{0}(980)} \times \varepsilon_{reco}^{B_{s}^{0} \rightarrow J/\psi \phi}}{N_{B_{s}^{0} 
\rightarrow J/\psi \phi} \times \varepsilon_{reco}^{B_{s}^{0} \rightarrow J/\psi f_{0}(980)} }.
\end{equation}
All that is required to measure a relative branching fraction are the relative yields and the relative reconstruction efficiencies of
the two decay modes,  $\varepsilon_{reco}^{B_{s}^{0} \rightarrow J/\psi \phi}$ 
and $\varepsilon_{reco}^{B_{s}^{0} \rightarrow J/\psi f_{0}(980)}$.

\section{Selection Cuts}

\subsection{Analysis Cuts} 

The data set of an integrated luminosity of approximately 8~fb$^{-1}$ was
divided into four periods corresponding to different detector configurations
called RunIIa, RunIIb1, RunIIb2 and RunIIb3.

The initial sample of
$B_{s}^{0} \rightarrow J/\psi f_{0}(980)$ was found by first reconstructing
$J/\psi \rightarrow \mu^{+} \mu^{-}$ candidates by requiring that two oppositely charged 
muon candidates with transverse momentum $p_{T}$ $>$ 1.5 GeV form a common vertex.
Since the D0 detector has a limited ability to separate kaons from pions, 
all reconstructed tracks not associated to a $J/\psi$ are considered for reconstructing $f_{0}(980)$ and $\phi$ candidates.  
The tracks are assigned the pion mass when searching for
 $B_{s}^{0} \rightarrow J/\psi f_{0}(980)$ and the kaon mass when searching for  $B_{s}^{0} \rightarrow J/\psi \phi$.
Two tracks with a minimum $p_{T}$ of 300 MeV, having an invariant mass 0.7 GeV $<$ $M_{\pi^{+} \pi^{-}}$ $<$ 1.2 GeV, 
and being consistent with coming from a common vertex were considered as $f_{0}(980)$ candidates. 
Finally, the $\mu^{+} \mu^{-} \pi^{+}\pi^{-}$ candidates were required to have a common vertex and have an invariant mass between 5.0--5.8 GeV.

Similar requirements were applied to the initial sample of $B_{s}^{0} \rightarrow J/\psi \phi$ candidates. The only different requirements were that 
0.91 GeV $<$ $M_{K^{+} K^{-}}$ $<$ 1.05 GeV and   the $\mu^{+} \mu^{-} K^{+}K^{-}$ candidates were required to have an invariant mass between 5.0--5.8 GeV.  Due to the invariant mass requirements on $M_{\pi^{+} \pi^{-}}$ and $M_{K^{+} K^{-}}$, two tracks cannot be considered 
both a $f_{0}(980)$ and a $\phi$ candidate.
The final data sample was then formed by applying the additional requirements:

\begin{itemize}
\item{All runs without optimal performance of muon, silicon microstrip and central fiber trackers are omitted .}
\item{All events that only fired a trigger that required muons with a large impact parameter were removed.}
\end{itemize}
$J/\psi$ selection:
\begin{itemize}
\item{Both muons are required to be detected as a track segment in either one or three layers of the muon system and be matched to a central track.}
\item{At least one muon must be detected as a track segment in three layers of the muon system.}
\item{Both muons must have at least one hit in the silicon microstrip tracker.}
\item{2.9 GeV $<$  $M_{\mu^{+} \mu^{-}}$ $<$ 3.2 GeV} 
\end{itemize}
$f_{0}(980)$ ($\phi$) selection:
\begin{itemize}
\item{Both pions (kaons) from the $f_{0}(980)$ ($\phi$) candidate must have at least 2 hits in the central fiber tracker.}
\item{Both pions (kaons) from the $f_{0}(980)$ ($\phi$) candidate must have at least 2 hits in the silicon microstrip tracker.}
\item{Both pions (kaons) from the $f_{0}(980)$ ($\phi$) candidate must have at least 8 hits total in the silicon microstrip tracker and the central fiber tracker.}
\item{The momentum of the leading pion (kaon) from the $f_{0}(980)$ ($\phi$) candidate must be greater than 1.4 GeV.}
\item{$f_{0}(980)$ ($\phi$) candidate $p_{T}$ must be greater than 1.6 GeV.}
\end{itemize}
$B_{s}^{0}$ selection:
\begin{itemize}
\item{0.91 GeV $<$ $M_{\pi^{+}\pi^{-}}$ $<$  1.05 GeV (when searching for $J/\psi f_{0}(980)$.)}
\item{1.01 GeV $<$ $M_{K^{+}K^{-}}$ $<$ 1.03 GeV (when searching for $J/\psi \phi$.)}
\item{$p_{T}(B_{s}^{0})$ $>$ 5.0 GeV }
\item{Proper decay length \cite{pdl}, $L$, significance, $L/\sigma(L)$ $>$ 5, where $\sigma(L)$ is the uncertainty on the proper decay length.} 
\end{itemize}
 
\subsection{Boosted Decision Trees}

\begin{figure}
\includegraphics[scale=0.6]{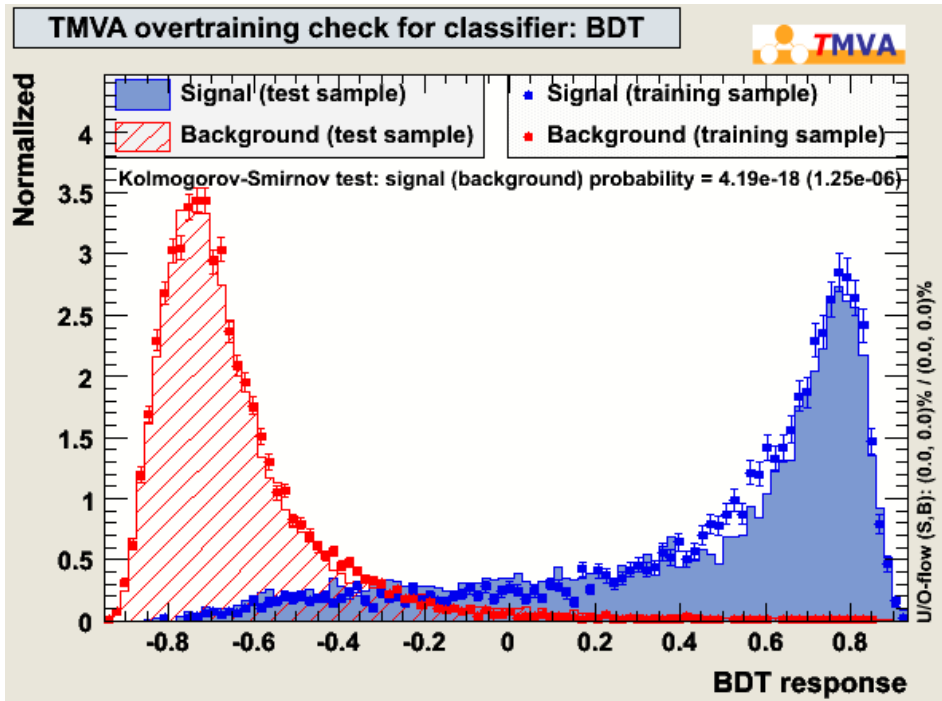}
\caption{\label{fig:BDT_latest} BDT distribution after training for both  signal (blue) and inclusive 
background (red).}
\end{figure}

\begin{figure}
\includegraphics[scale=0.6]{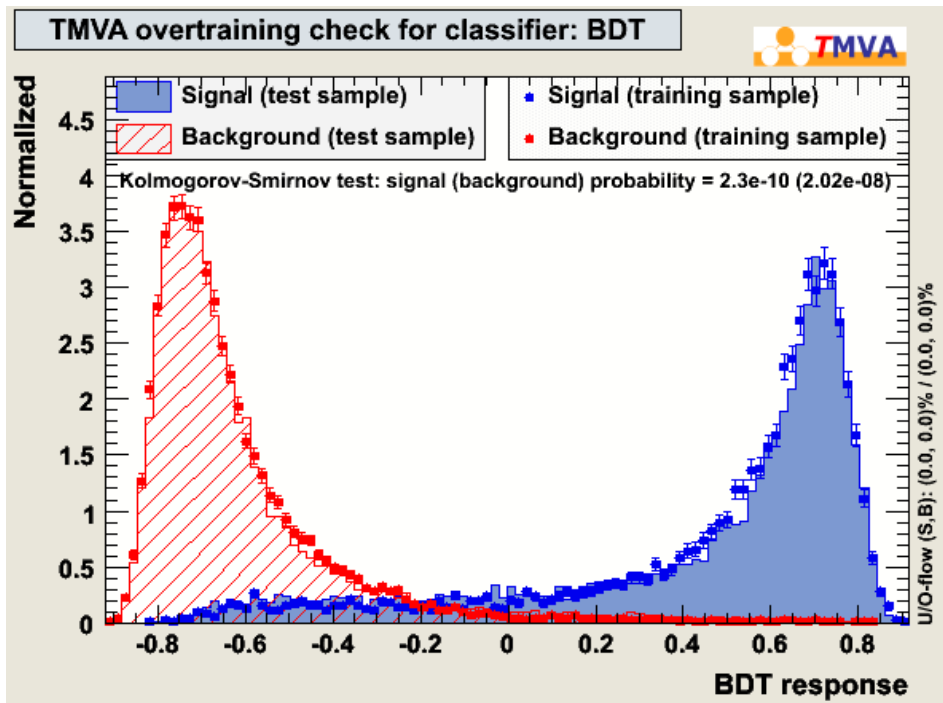}
\caption{\label{fig:BDT_latest_pro} BDT distribution after training for both signal (blue) and prompt background
(red).}
\end{figure}

It is known that boosted decision trees (BDT) \cite{BDT1,BDT2} are a powerful tool for separating signal from background.
Signal and background samples are used to train the BDT and a discriminant is determined for each event.
By making a selection on the value of the BDT discriminant, the signal to background ratio can be vastly improved.
We use the Monte Carlo (MC) {\sc pythia} program \cite{pythia} to generate $B_{s}^{0}$ and the {\sc evtgen} program \cite{Evtgen} to
simulate its decay.  
Two MC background samples were produced:
a prompt sample (directly produced $J/\psi$) and an inclusive sample (all decay processes
$B_{s}^{0}\rightarrow J/\psi + X$).  A MC signal sample of $B_{s}^{0}\rightarrow J/\psi f_{0}(980)$ events was then used to train
the BDT on both the prompt and inclusive background. 
A BDT discriminant was found for both the prompt and 
inclusive sample and used in the analysis.  A total of 36 different kinematic variables were used to train the BDT
consisting of isolation variables, transverse momentum of the daughters and grand-daughters of the $B_{s}^{0}$ and vertex 
quality of the $B_{s}^{0}$ and its daughters.   
Figures \ref{fig:BDT_latest} and \ref{fig:BDT_latest_pro} show the BDT distributions for the training and test samples 
for the inclusive and prompt background. 

The BDT cuts were determined only using the 1~fb$^{-1}$ of RunIIa data. 
A narrow window around the nominal $f_{0}(980)$
mass was chosen to keep the signal to noise ratio high.
Using a mass cut of 0.96--1.0 GeV on the $\pi^{+}\pi^{-}$ mass, the BDT cut value was chosen where both
S/$\sqrt{B}$ and the signal yield were high.  In this way, the BDT discriminant for both the inclusive and prompt BDT was required to be greater than 0.35.

\section{Yield Results}

\begin{figure}
\includegraphics[scale=0.5]{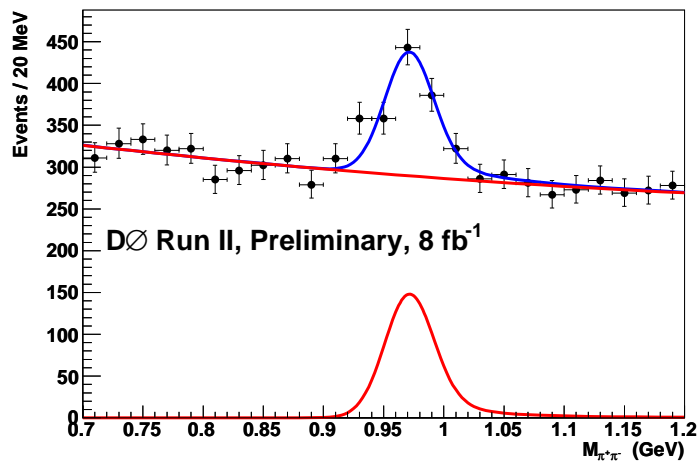}
\caption{\label{fig:f0_mass} $\pi^{+}\pi^{-}$ invariant mass distribution peaking at the $f_{0}(980)$ mass when the $J/\psi \pi^{+}\pi^{-}$ mass is
$\pm 2\sigma$ around the fitted $B_{s}^{0}$ mass.}
\end{figure}

\begin{figure}
\includegraphics[scale=0.5]{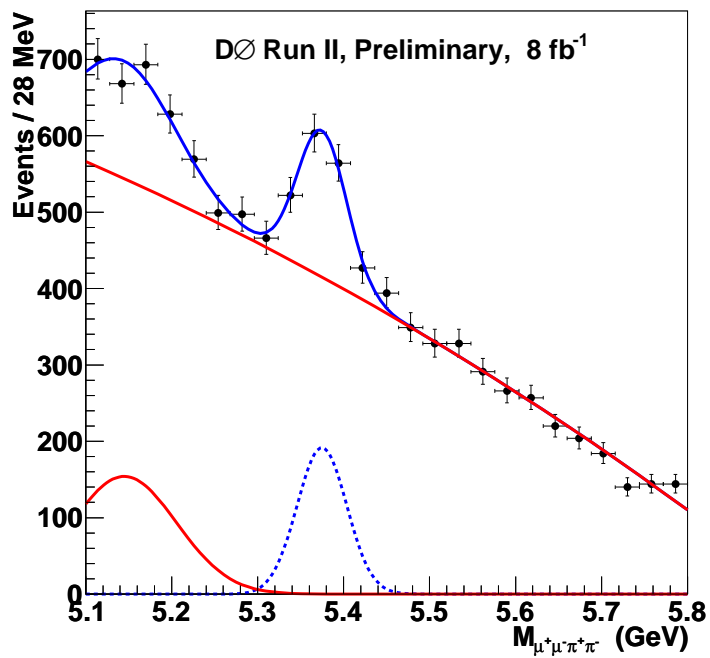}
\caption{\label{fig:jpsi_f0_mass} $\mu^{+} \mu^{-} \pi^{+}\pi^{-}$ mass distribution peaking at the $B_{s}^{0}$ mass when the
 $\pi^{+}\pi^{-}$ mass is between 0.91 and 1.05 GeV}
\end{figure}

A clear $B_{s}^{0}$ peak is found when the $\pi^{+}\pi^{-}$ invariant mass is near
the nominal $f_{0}(980)$ mass.  It is expected that the $B_{s}^{0}$ signal can be fitted to a Gaussian distribution, which provides a fitted
mean mass ($\mu$) and width ($\sigma$) for the $B_{s}^{0}$ peak.  Since backgrounds are large, a cut of $\pm 2\sigma$ around
the fitted $B_{s}^{0}$ peak is used to identify the $f_{0}(980)$ mass peak.  
A clear $f_{0}(980)$ mass peak is observed when the $\mu^{+} \mu^{-} \pi^{+}\pi^{-}$ invariant mass is within $\pm 2\sigma$ of the fitted $B_{s}^{0}$ mass,
see Fig. \ref{fig:f0_mass}. 
To decide on a $\pi^{+}\pi^{-}$ mass window to use for this analysis, a fit to the $f_{0}(980)$ mass peak is performed. 
The $f_{0}(980)$ has a large width \cite{PDG} and is just under the $KK$ mass threshold.  This changes the line shape from
a simple Breit Wigner form, particularly for higher masses and so   
the $\pi^{+}\pi^{-}$ mass distribution is fitted using a functional form based on Flatt\'e \cite{Flatte}, convoluted with a Gaussian function,  
that takes into account the opening of the $KK$ threshold.
The lineshape found from fitting the $f_{0}(980)$ in MC is used to fit the data. 
A  $\pi^{+}\pi^{-}$ invariant mass cut of 0.91--1.05 GeV is applied to identify $B_{s}^{0}\rightarrow J/\psi f_{0}(980)$ and is shown in Fig. \ref{fig:jpsi_f0_mass}.  The  $B_{s}^{0} \rightarrow J/\psi f_{0}(980)$ mass distribution was fit to a Gaussian signal with a background function consisting
of a second-degree polynomial and a Gaussian function at lower invariant mass to take into account partially reconstructed $B$ decays.

\begin{figure}
\includegraphics[scale=0.5]{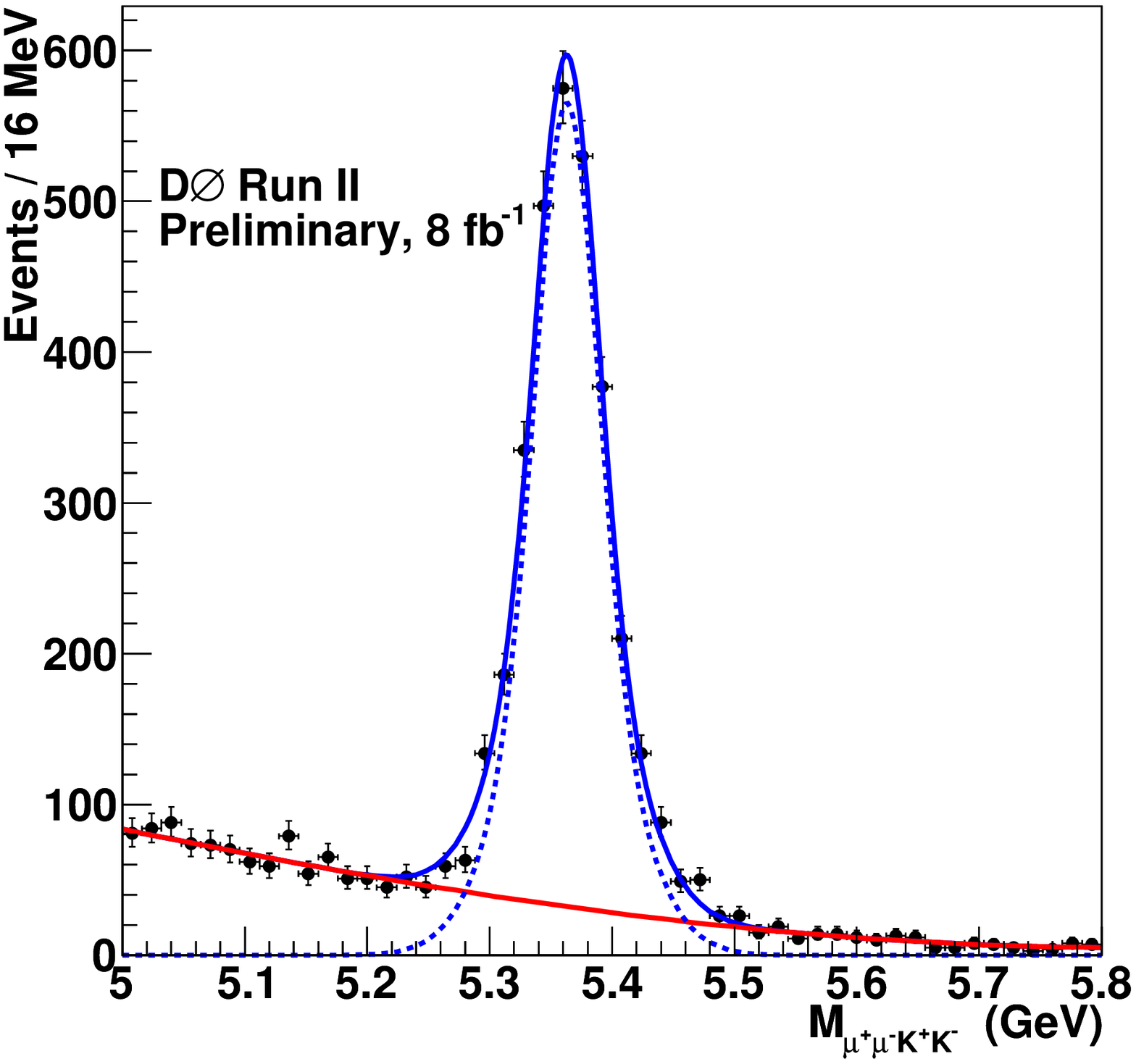}
\caption{\label{fig:jpsi_phi_mass} $\mu^{+} \mu^{-} K^{+} K^{-}$ mass distribution peaking at the $B_{s}^{0}$ mass from 
8~$\rm{fb}^{-1}$ of data}
\end{figure}

Using identical cuts (except for the cut on the $\phi$ mass), a clear $J/\psi \phi$ peak is found
and is shown in Fig. \ref{fig:jpsi_phi_mass}.  Since the $\phi$ peak is so narrow, the backgrounds are much smaller for 
$B_{s}^{0}\rightarrow J/\psi \phi$.

An unbinned likelihood fit was used to determine the candidate yields in each sample.  
 The fit to the $J/\psi f_{0}(980)$ mass distribution shown in Fig. \ref{fig:jpsi_f0_mass} gives the following results (statistical
uncertainties only):
\begin{eqnarray*}
 B_{s}^{0} \; {\rm mass} = 5.3747 \pm 0.0036 \; {\rm GeV;} \;\;
 \sigma = 0.0290 \pm 0.0044 \; {\rm GeV};  \; \;
498 \pm 74 \; B_{s}^{0} \rightarrow J/\psi f_{0}(980) \; {\rm candidates.}
\end{eqnarray*}

The $\mu^+\mu^-K^+K^-$ mass distribution was fit for a  $B_{s}^{0} \rightarrow J/\psi \phi$ signal
using a double Gaussian function with a second-order polynomial background.
A fit to the $J/\psi \phi$ distribution shown in Fig. \ref{fig:jpsi_phi_mass} gives the following results (statistical uncertainties only):
\begin{eqnarray*}
 B_{s}^{0} \; {\rm mass} = 5.3631 \pm 0.0008 \; {\rm GeV;} \;\;  2863 \pm 61 \; B_{s}^{0} \rightarrow J/\psi \phi \; {\rm candidates.}
\end{eqnarray*}

\section{Efficiencies}

To determine the efficiencies of the analysis, MC signal samples were used.  To take into account the effects of the instantaneous
luminosity, the MC samples were overlaid with zero bias data collected during each run period.
In the generation of both the $J/\psi \phi$ and the $J/\psi f_{0}(980)$ signal MC's, a preselection requirement of $p_{T}$
$>$ 0.4 GeV was demanded on both kaons (pions) from the $\phi$($f_{0}(980)$).  Since the $p_{T}$ distributions
for the pions and kaons may be different, the preselection efficiencies of this cut must be determined.
To determine the preselection cut efficiencies, two additional MC sets were also generated with no
$p_{T}$ cuts on the pions (kaons).  By comparing these two results, the preselection cut efficiencies were determined.

We found that the reconstruction efficiencies depended heavily on the MC sample used since the instantaneous luminosity
was different for the various run periods, therefore we determined 
the reconstruction efficiencies for each run range separately.  The instantaneous luminosities for runs taken during 
RunIIb3 were similar to the instantaneous luminosities for
runs taken during RunIIb2 so the reconstruction efficiencies found from RunIIb2 were used for RunIIb3.
Table \ref{tab:effic} shows the results on the efficiency analysis using MC signal samples.
Table \ref{tab:effic} shows that the absolute reconstruction efficiencies vary in each run period, however Table \ref{tab:ranges} show
that the relative reconstruction efficiencies are relatively stable.  However, the differences in the relative reconstruction efficiency
is considered a systematic uncertainty on $R$.

\begin{table}
\begin{center}
\caption{The reconstruction efficiency 
for $B_{s}^{0} \rightarrow J/\psi \phi$ and $B_{s}^{0} \rightarrow J/\psi f_{0}(980)$ for various running periods.}
\begin{tabular}{|c|c|}
\hline
Sample & total reconstruction efficiency \\
\hline
$B_{s}^{0} \rightarrow J/\psi \phi$ RunIIa & 0.0231 $\pm$ 0.0004\\
$B_{s}^{0} \rightarrow J/\psi \phi$ RunIIb1 & 0.0191 $\pm$ 0.0004\\
$B_{s}^{0} \rightarrow J/\psi \phi$ RunIIb2 & 0.00636 $\pm$ 0.00018\\
$B_{s}^{0} \rightarrow J/\psi f_{0}(980)$ RunIIa & 0.0191 $\pm$ 0.0004\\
$B_{s}^{0} \rightarrow J/\psi f_{0}(980)$ RunIIb1 & 0.0146 $\pm$ 0.0003\\
$B_{s}^{0} \rightarrow J/\psi f_{0}(980)$ RunIIb2 & 0.00529 $\pm$ 0.00015\\
\hline
\end{tabular}
\label{tab:effic}
\end{center}
\end{table}

\begin{table}
\caption{\label{tab:ranges} Reconstruction efficiencies for different run periods.}
\begin{center}
\begin{tabular}{|c|c|}

\hline
Run period & Relative reconstruction efficiency $ \frac { \varepsilon_{reco}^{B_{s}^{0} \rightarrow J/\psi \phi} } {\varepsilon_{reco}^{B_{s}^{0} \rightarrow J/\psi f_{0}(980)}} $ \\
\hline
RunIIa  & 1.21 $\pm$ 0.03 \\
RunIIb1 & 1.31 $\pm$ 0.04 \\
RunIIb2 & 1.20 $\pm$ 0.05 \\
\hline
\end{tabular}
\end{center}
\end{table}

%\section{Ratio of Branching Fractions}

%Using the relative yields and the relative efficiencies shown above we find for the ratio of branching fractions:
%$$ R = 
%\frac{N_{B_{s}^{0} \rightarrow J/\psi f_{0}(980)} \times \varepsilon_{reco}^{B_{s}^{0} \rightarrow J/\psi \phi} } {N_{B_{s}^{0} \rightarrow J/\psi} \phi \times
%\varepsilon_{reco}^{B_{s}^{0} \rightarrow J/\psi f_{0}(980)}} = 0.210 \pm 0.032\thinspace({\rm stat}) $$

\section{Systematic Uncertainty Studies}

\subsection{$B_{s}^{0} \rightarrow J/\psi \pi^{+} \pi^{-}$ background studies}

One possible peaking background that affects the $B_{s}^{0} \rightarrow J/\psi f_{0}(980)$ yield 
measurement is the non-resonant $B_{s}^{0} \rightarrow J/\psi \pi^{+} \pi^{-}$ background.  
This background was studied by measuring the $B_{s}^{0}$ yields in $\pi^{+} \pi^{-}$ invariant mass less
than the $f_{0}(980)$ mass.  The $\pi^{+} \pi^{-}$ mass distribution from $B_{s}^{0} \rightarrow J/\psi 
\pi^{+} \pi^{-}$ background where the $\pi^{+} \pi^{-}$ are non-resonant  
should have a much broader distribution, so determining the $B_{s}^{0}$ yield
for lower $\pi^{+} \pi^{-}$ masses will allow a determination of the contamination in
the $f_{0}(980)$ signal region.

%\begin{table}
%\caption{\label{tab:pipiyields}Yield of events in $B_{s}^{0} \rightarrow J/\psi 
%\pi^{+} \pi^{-}$ non-resonant MC for two different $\pi^{+}\pi^{-}$ mass regions.}
%\begin{center}
%\begin{tabular}{|c|c|}

%\hline
%$\pi^{+} \pi^{-}$ mass region & MC yield \\
%\hline
%Below $f_{0}(980)$ mass (0.80--0.90 GeV) & 637 $\pm$ 26 \\
%$f_{0}(980)$ Signal region (0.91--1.05 GeV) & 979 $\pm$ 33 \\
%\hline
%\end{tabular}
%\end{center}
%\end{table}

In determining the $\pi^{+} \pi^{-}$ mass window to study, it is important to choose a window where one does not
expect other resonances (i.e., $B_{s}^{0} \rightarrow J/\psi K^{*}$).  The  $\pi^{+} \pi^{-}$ mass window of 0.8--0.9 GeV
was chosen since in this mass range there should not be any $B_{s}^{0} \rightarrow J/\psi K^{*}$ events.  
In fitting the distribution for any possible signal, the signal $\mu$ and $\sigma$ 
are constrained to be the values found from the fit to the $B_{s}^{0}$ mass in the $f_{0}(980)$ signal region. 
The fit yields 80 $\pm$ 75 events, giving no statistically significant evidence 
of any $B_{s}^{0} \rightarrow J/\psi \pi^{+} \pi^{-}$ non-resonant background, so no correction was applied.

\subsection{Analysis cut variation}

To cross check that the results do not vary with the exact value of the analysis cuts, 
the choice for each analysis cut was varied around its nominal
value.  This is an important test since the selection criteria was determined with 1~fb$^{-1}$
data from RunIIa, and it is important to verify that this did not introduce a bias into the measurement.
Table \ref{tab:variation} shows the results from this study.
As can been seen from the table, the value of $R$ does not depend significantly on the exact choice of
selection requirement.  

\begin{table}
\caption{\label{tab:variation} Fractional change due to varying the exact choice of analysis cuts
on the relative branching fraction}
\begin{center}
\begin{tabular}{|c|c|c|c|c|c|}

\hline
Cut & $\varepsilon$ ($J/\psi \phi)$ & $\varepsilon$ ($J/\psi f_{0}$) & event yield $B_{s}^{0}\rightarrow J/\psi \phi$ & event yield $B_{s}^{0}\rightarrow J/\psi f_{0}$ & effect on $R$ \\
\hline
BDT inc $>$ 0.3 & 1.000 & 1.017 & 1.020  & 0.958 & 0.96 \\
BDT inc $>$ 0.4 & 0.993 & 0.980 & 0.975 & 0.945 & 0.98 \\ 
BDT pro $>$ 0.3 & 1.000 & 1.002 & 1.000   & 1.007 & 1.01 \\
BDT pro $>$ 0.4 & 1.002 & 1.000 & 1.000   & 0.991 & 0.99  \\
$p_T(B_{s}^{0}) >$~4.5 GeV & 0.997 & 1.000 & 1.000 & 1.000 & 0.99 \\
$p_T(B_{s}^{0}) >$~5.5 GeV & 1.000 & 0.995 & 1.000 & 0.952 & 0.95 \\
$p_T(f_{0}(980))>$~1.0 GeV & 1.000 & 1.000 & 1.000 & 1.000 & 1.00 \\
$p_T(f_{0}(980))>$~2.0 GeV & 1.000 & 0.987 & 1.000 & 0.980 & 0.99  \\
$\pi$/K $p_{T}$ $>$ 1.0 GeV & 1.210 & 1.099 & 1.172 & 1.133 & 1.06 \\
$\pi$/K $p_{T}$ $>$ 1.8 GeV & 0.724 & 0.771 & 0.797 & 0.744 & 0.88 \\
$L/\sigma(L)$ $>$ 4 & 1.057 & 1.047 & 1.056 & 1.035 & 1.01 \\
$L/\sigma(L)$ $>$ 6 & 0.946 & 0.951 & 0.944  & 0.967  & 1.02 \\
\hline
\end{tabular}
\end{center}
\end{table}

\subsection{Fitting cross checks}

Due to large backgrounds arising from combinatorics and partially reconstructed $B$ decays, 
there are significant uncertainties in the exact background shape.  
Therefore different parameterizations were used to
describe the background and different fit regions were used to fit the data.
The background polynomial was changed from a second-degree polynomial to a third-degree polynomial.
The fit range was changed from the nominal 5.1--5.8 GeV and finally a different functional form for the background
was used by changing the background shape to a polynomial plus an exponential.  

%Results are summarized in Table \ref{tab:fitsys}.
\begin{table}
\caption{\label{tab:fitsys} Effects of changing the fitting choices}
\begin{center}
\begin{tabular}{|c|c|}

\hline
Parameter & $B_{s}^{0}\rightarrow J/\psi f_{0}(980)$ yield \\
\hline
Nominal fit (Gaussian signal + second order polynomial background with fit range 5.1--5.8 GeV) & 498 $\pm$ 74 \\
Third degree polynomial background & 446 $\pm$ 72 \\
Background function exponential+polynomial & 423 $\pm$ 67 \\ 
Fit range 5.1--5.6 GeV & 437 $\pm$ 78 \\ 
Fit range 5.15--5.8 GeV  & 427 $\pm$ 63 \\
Fit range 5.05--5.8 GeV & 449 $\pm$ 71 \\ 
\hline 
\end{tabular}
\end{center}
\end{table}

As can be seen from Table \ref{tab:fitsys}, there is a fairly large variation in the number 
of signal events for $B_{s}^{0}\rightarrow J/\psi f_{0}(980)$, indicating
that the background shape is difficult to model. This fitting systematic gives the largest systematic uncertainty on $R$.
A study was performed using same-sign pions and forming the mass distribution from
$\mu^{+} \mu^{-} \pi^{\pm} \pi^{\pm}$.  However, it was found the the same sign pion distribution did not describe the
measured background and so could not be used to help constrain the background shape.
A similar study of varying the fitting choices was performed on the $B_{s}^{0}\rightarrow J/\psi \phi$ sample, 
however since the backgrounds are much smaller and
easier to describe the event yield numbers changed by less than 1\%.  

%The presence of $B^0 \rightarrow J/\Psi \pi^+ \pi^-$ was checked by including this channel
%in the fit.  A fit consistent with zero events was found, although forcing a number of $B^0$ events while still maintaining an
%acceptable fit resulted in a variation of yield within the indicated systematic uncertainty.

A summary of the uncertainties on the BR are summarized in Table \ref{tab:unc}.

\begin{table}
\caption{\label{tab:unc} Statistical and systematic uncertainties in branching fraction ratio, $R$}
\begin{center}
\begin{tabular}{|c|c|}

\hline
Source & Uncertainty \\
\hline
Statistical & 0.149   \\
Systematic from fitting & 0.150\\
Systematic from different MC samples & 0.0858 \\ 
\hline
\end{tabular}
\end{center}
\end{table}

\section{Final branching fraction ratio}

The decay $B_{s}^{0}\rightarrow J/\psi f_{0}(980)$ is an interesting decay mode since it can allow
a measurement of the CP-violating phase in $B_{s}^{0}$ mixing.  

A measurement of the relative branching fraction using approximately 8~fb$^{-1}$ of data yields:

$$ R =  \frac{ \mathcal{B}(B_{s}^{0}\rightarrow J/\psi f_{0}(980)    ;f_{0}(980)    \rightarrow \pi^{+} \pi^{-})  }
        { \mathcal{B}(B_{s}^{0}\rightarrow J/\psi \phi;\phi \rightarrow K^{+} K^{-}    )} = 0.210 \pm 0.032\thinspace(\rm{stat}) \pm 0.036\thinspace (\rm{syst}). $$

The relative branching fraction of  $B_{s}^{0} \rightarrow J/\psi f_{0}(980), 
f_{0}(980) \rightarrow \pi^{+}\pi^{-}$ to $B_{s}^{0} \rightarrow J/\psi \phi, \phi \rightarrow K^{+}K^{-}$ should be large enough to allow a measurement of
the CP-viiolating phase in $B_{s}^{0}$ mixin using the decay  $B_{s}^{0} \rightarrow J/\psi f_{0}(980)$.  An analysis to measure $\phi_{s}$ using the 
decay $B_{s}^{0} \rightarrow J/\psi f_{0}(980)$ is currently being pursued.

\bigskip % extra skip inserted

\begin{thebibliography}{99}
\bibitem{pdl} The proper decay length is defined as $L_{xy}(M_{B_{s}^{0}}/p_{T})$, where $p_{T}$ is the transverse momentum of the $B_{s}^{0}$, 
$M_{B_{s}^{0}}$ is the world average mass of the $B_{s}^{0}$, and $L_{xy}$ is the transverse distance between the primary vertex 
and the four track vertex of the $B_{s}^{0}$ candidate. Primary vertices are determined by using the beamspot as a contraint and finding
the vertex which contains the most tracks.  Then using the other tracks not associated with the first primary vertex, but still beam
constrained, search for a second primary vertex.  This process is then repeated until no additional primary vertices are found.
If there is more than one primary vertex in an event, the primary vertex nearest the $J/\Psi$ candidate is selected.
\end{thebibliography}

\end{document}